\documentclass[%
reprint,
superscriptaddress,
amsmath,amssymb,
aip,
jcp,
citeautoscript,
]{revtex4-2}
\usepackage{graphicx}
\usepackage{dcolumn}
\usepackage{bm}

\usepackage{siunitx}
\usepackage{mathrsfs,amsmath}
\usepackage{mathptmx}
\usepackage{etoolbox}
\usepackage{lipsum}
\usepackage{mathtools}
\usepackage{fontenc}
\usepackage{xcolor}

\begin{document}
\title{Measurement principles for quantum spectroscopy of molecular materials with entangled photons}

\author{Luca~Moretti}
\affiliation{Dipartimento di Fisica, Politecnico di Milano, Piazza Leonardo da Vinci 32, Milano, Italy}
\affiliation{Department of Physics and Center for Functional Materials, Wake Forest University, 1834 Wake Forest Road, Winston-Salem, NC~27109, United~States}

\author{Esteban Rojas-Gatjens}
\affiliation{Department of Physics and Center for Functional Materials, Wake Forest University, 1834 Wake Forest Road, Winston-Salem, NC~27109, United~States}
\affiliation{School of Chemistry and Biochemistry, Georgia Institute of Technology, Atlanta, USA.}

\author{Lorenzo Uboldi}
\affiliation{Department of Physics and Center for Functional Materials, Wake Forest University, 1834 Wake Forest Road, Winston-Salem, NC~27109, United~States}
\affiliation{Dipartimento di Fisica, Politecnico di Milano, Piazza Leonardo da Vinci 32, Milano, Italy}

\author{David Otto Tiede}
\affiliation{Department of Physics and Center for Functional Materials, Wake Forest University, 1834 Wake Forest Road, Winston-Salem, NC~27109, United~States}
\affiliation{Institute of Materials Science of Sevilla, Spanish National Research Council, Américo Vespucio, 49. 41092 – Sevilla}

\author{Evan J Kumar}
\affiliation{Department of Physics and Center for Functional Materials, Wake Forest University, 1834 Wake Forest Road, Winston-Salem, NC~27109, United~States}

\author{Chiara Trovatello}
\affiliation{Dipartimento di Fisica, Politecnico di Milano, Piazza Leonardo da Vinci 32, Milano, Italy}
\affiliation{Department of Mechanical Engineering, Columbia University, New York, 10027, NY, USA}

\author{Fabrizio Preda}
\affiliation{NIREOS S.R.L., Via G. Durando 39, 20158 Milano, Italy}

\author{Antonio Perri}
\affiliation{NIREOS S.R.L., Via G. Durando 39, 20158 Milano, Italy}

\author{Cristian~Manzoni}
\affiliation{IFN-CNR, Dipartimento di Fisica, Politecnico di Milano, Piazza Leonardo da Vinci 32, Milano, Italy}

\author{Giulio~Cerullo}
\affiliation{Dipartimento di Fisica, Politecnico di Milano, Piazza Leonardo da Vinci 32, Milano, Italy}
\affiliation{IFN-CNR, Dipartimento di Fisica, Politecnico di Milano, Piazza Leonardo da Vinci 32, Milano, Italy}

\author{Ajay~Ram~Srimath~Kandada}
\affiliation{Department of Physics and Center for Functional Materials, Wake Forest University, 1834 Wake Forest Road, Winston-Salem, NC~27109, United~States}
\email{srimatar@wfu.edu}

\date{\today}

\begin{abstract}
Nonlinear spectroscopy with quantum entangled photons is an emerging field of research which holds the promise to achieve superior signal to noise ratio and effectively isolate many-body interactions. Photon sources used for this purpose however lack the frequency tunability and spectral bandwidth demanded by contemporary molecular materials. Here, we present design strategies for efficient spontaneous parametric downconversion to generate biphoton states with adequate spectral bandwidth and at visible wavelengths. Importantly, we demonstrate, by suitable design of the nonlinear optical interaction, the scope to engineer the degree of spectral correlations between the photons of the pair. We also present an experimental methodology to effectively characterize such spectral correlations. Importantly, we believe that such a characterization tool can be effectively adapted as a spectroscopy platform to optically probe system--bath interactions in materials. 

\end{abstract}
\maketitle

\newpage
\section{\label{sec:intro}Introduction}

Ultrafast optical spectroscopy techniques that exploit the high temporal resolution enabled by femtosecond lasers have played a pivotal role in unveiling early snapshots of many critical photoinduced processes, in fields ranging from biochemistry to materials science. These include charge separation mechanisms that drive next-generation solar cells~\cite{falke2014coherent, gelinas2014ultrafast, grancini2013hot}, rate limiting steps of photo-chemical reactions in photosynthetic reaction centers~\cite{dahlberg2017mapping, cao2020quantum} and retinal chromophores~\cite{polli2010conical}, and quantum dynamics in condensed phase~\cite{chen2021ultrafast, de2012quantum, micha2007quantum, lloyd20212021}. These successes, combined with rapidly increasing accessibility of the required instrumentation, have resulted in the ubiquity of ultrafast spectroscopy techniques in physical, chemical and biological research; however, challenges still exist that limit their scope. 

The most conspicuous of these challenges is their relatively low sensitivity, that prevents exploration of photophysical phenomena at low excitation densities~\cite{kanal2014100, preda2016broadband}. This is pertinent in certain scenarios, for example in the observation of energy conversion dynamics in photovoltaic materials that operate at relatively low solar intensities or when performing nonlinear microscopy of biological systems that degrade at high light intensities~\cite{tsai2006optical}. The ability to perform experiments at low photon flux becomes even more critical when ultrafast techniques are used to decipher many-particle signatures~\cite{meier1996femtosecond, mukamel2016communication, li2022optical}, quantum coherent phenomena~\cite{collini2013spectroscopic, cao2020quantum} or quasi-particle entanglement. In such scenarios, exploitation of the quantum mechanical properties of light has been suggested to offer unique advantages over the classical methodology of measuring amplitude and phase of the electromagnetic field of the optical probes~\cite{dorfman2016nonlinear, georgiades1995nonclassical, gea1989two, lee2006entangled, harpham2009thiophene, szoke2020entangled}.

One of the suggested methods to achieve quantum optical probes of ultrafast processes in molecular materials is through the use of quantum entangled photons~\cite{mukamel2020roadmap, kowalewski2017manipulating, schlawin2013two, dorfman2014stimulated, cuevas2018first, kalashnikov2017quantum}. Of particular relevance here are photons that are spectrally entangled~\cite{dorfman2021multidimensional, dayan2004two, oka2010efficient, fei1997entanglement} and thus effective in addressing various excited states and their interactions~\cite{kojima2004entangled, asban2021distinguishability, asban2021interferometric, raymer2021entangled, raymer2013entangled}.
The optical spectra of molecular materials of interest for optoelectronic applications exhibit relatively broad features, owing to inherently large homogeneous and inhomogeneous spectral broadening. Thus, to effectively probe these material systems, quantum spectroscopies must be equipped with sources of entangled photons with substantial spectral bandwidths. More importantly, the energies of these photons must match with the electronic transition energies which, for the molecular and solid-state systems most relevant for applications, correspond to the visible and near-infrared wavelengths.

Spectrally entangled photon pairs are often produced with high brightness through spontaneous parametric down-conversion (SPDC) in crystals with strong second-order ($\chi^{(2)}$) nonlinearity~\cite{joobeur1996coherence}. The design methodology of the nonlinear materials for the SPDC process is widely established, and there are indeed off-the-shelf crystals now available in the market. Notably, though, most of the SPDC photons are generated by either the fundamental or the second harmonic of a Ti:sapphire laser, and at photon energies of about 0.8\,eV (1550\,nm) or 1.55\,eV (800\,nm), and they do not offer large spectral bandwidths. This limited tunability and bandwidth substantially restricts the applicability of these crystal designs to molecular spectroscopy.

Here, we address this challenge by designing and experimentally implementing broadband entangled photons at visible wavelengths. We also provide an experimental methodology for the characterization of their joint spectral properties in view of their applications to spectroscopy. The manuscript is organized as follows: In Section~\ref{Sec:Calculations}, we first review the nonlinear optics of the SPDC process and highlight the similarities between its classical and quantum mechanical treatments. We discuss how phase-matching conditions, which are at the core of optical parametric amplifiers (OPAs), determine the joint spectral amplitude (JSA) of the quantum entangled photon state and accordingly the measure of entanglement. Based on these fundamental principles, we present and discuss designs of nonlinear crystals and the quantum descriptors of the resultant SPDC photon state. In Section~\ref{Sec:Experiment}, we present our perspective of using spectral correlations in a biphoton state as a material probe and plausible experimental schemes that enable them. We present an experimental implementation of one such scheme to estimate the spectral correlation in a biphoton state generated in a prototypical and custom-designed nonlinear crystal.

\section{Design principles of entangled photon generation}\label{Sec:Calculations}

\subsection{Spontaneous parametric downconversion}

\noindent In a semi-classical treatment of light-matter interaction, nonlinear optical processes originate from the nonlinear polarization in the material induced by a propagating intense light field.~\cite{boyd2020nonlinear, shih2020introduction}
Here we will limit our discussion to second-order nonlinear processes, for which two field interactions in a non-centrosymmetric medium result in a non-linear polarization that emits radiation at novel photon energies that are parametrically related to the input photon energies. For example, two input electric fields can result in an output whose frequency is the sum of the frequencies of the input fields. In the opposite process an intense input field is parametrically downconverted to two output fields at lower frequencies. This is the basis of OPA and SPDC. The latter is a powerful source of entangled photon pairs, as will be discussed in the following. 

In SPDC, a pump photon at energy $\hbar\omega_{p}$ is spontaneously downconverted in a nonlinear crystal into two photons, referred to as signal ($\hbar\omega_{s}$) and idler ($\hbar\omega_{i}$) photons. Energy conservation dictates that the sum of signal and idler photon energies is equal to the energy of the pump photon:
\begin{equation}
    \hbar\omega_s+\hbar\omega_i=\hbar\omega_p
    \label{Eq: SPDC1}
\end{equation}

The generation of the signal and idler photons can be either \emph{stimulated} or \emph{spontaneous}. In the first case, a seed field at frequency $\omega_s$ is injected together with the pump in the nonlinear medium, triggering the emission of a signal photon at the same frequency  $\omega_s$ and of the corresponding idler. This is exploited in OPAs, which are work-horse systems to generate tunable ultrashort optical pulses by transferring energy from a pump beam to a low-power signal beam. SPDC, on the other hand, occurs when only the pump beam interacts with the nonlinear medium, and a signal photon is emitted from a virtual level at a potentially random frequency $\omega_s$, together with the corresponding idler photon $\omega_i$. This process can be also thought of as parametric amplification of the vacuum quantum fluctuations~\cite{manzoni2009optical}, a mechanism also known as optical parametric generation. The unseeded OPA is essentially a source of photon pairs at $\omega_s$ and $\omega_i$ with entangled frequency, emission time, and/or polarization. 
Note that the emitted photons are labeled following the convention of the seeded OPA in which the signal corresponds to the seed frequency. In the case of the SPDC discussed here, the emitted photons are equivalent, hence in the following they will be arbitrarily labeled signal and idler.

\subsection{Quantum mechanical treatment of SPDC}

\begin{figure*}
    \centering
    \includegraphics[width=15 cm]{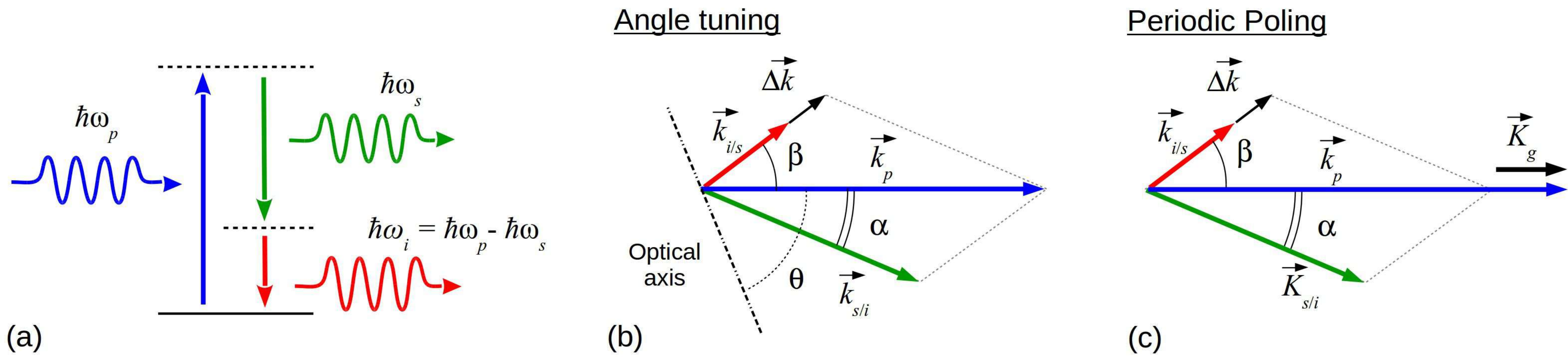}
    \caption{Scheme of parametric amplification. (a) Corpuscular interpretation of the interaction: after the absorption of one pump photon, two photons (signal and idler) are released, at frequencies that preserve the overall energy. The dashed lines are virtual levels, which only exist during the nonlinear interaction. (b) Wavevectors of signal, idler and pump photons in a birefringent crystal. The panel also shows the $\Delta \vec{k}$ vector and the optical axis of the crystal. (c) The same as (b), in a periodically poled crystal.}
    \label{fig1_PA}
\end{figure*}

\noindent Quantum mechanically, the SPDC process in a nonlinear crystal can be described by the time-dependent interaction Hamiltonian~\cite{grice1997spectral, keller1997theory}:

\begin{equation}
    H_I(t) = \int_r d^3r \chi^{(2)}\hat{E_p}^{(+)}\hat{E_s}^{(-)}\hat{E_i}^{(-)} + c.c,\label{Eq:HI}
\end{equation}

\noindent where $\chi^{(2)}$ is the second order nonlinear susceptibility of the nonlinear crystal and $\hat{E_j}^{(+)}$ are electric field operators which can be described in terms of the photon operators $\hat{a}_j(\omega_j)$,  which destroy a photon at frequency $\omega_j$, as:

\begin{equation}
    \hat{E_j}^{(+)} = \int d\omega_j A(\omega_j)\hat{a_j}(\omega_j) \exp i[k_j(\omega_j)z-\omega_jt].
\end{equation}

\noindent In the equation above, $A(\omega_j)$ is the amplitude function related to the photon frequency and wavevector, and $\hat{E_j}^{(-)}$ is the Hermitian conjugate of $\hat{E_j}^{(+)}$, which is the field operator associated with the photon creation. The index $j$ runs over the pump, signal, and idler photons. Since the nonlinear processes happen only at high pump intensities, the corresponding field operator can be replaced by a classical field, given by $\tilde{\Gamma} (t) \exp (ik_p(\omega_p)z)$. Time evolution of the photon state vector can now be derived using Eq.~\ref{Eq:HI} and integrating it over the finite duration of the pump pulse and the length L of the crystal, obtaining the expression:
\begin{equation}
    |\psi\rangle = \frac{2\pi A}{i\hbar}\int \int d\omega_id\omega_s \mathcal{F}(\omega_i,\omega_s) \hat{a_i}^{\dagger}(\omega_i) \hat{a_s}^{\dagger}(\omega_s)|0\rangle
    \label{Eq:biphoton}
\end{equation}

\noindent where $|0\rangle$ is the vacuum state, representing the absence of the biphoton amplitude and $\mathcal{F} (\omega_i,\omega_s)$ is the biphoton field amplitude, also known as joint spectral amplitude (JSA), that can be written as:
\begin{equation}
    \mathcal{F}(\omega_i,\omega_s) = \Gamma (\omega_s+\omega_i) \Phi(\omega_s,\omega_i)\label{Eq:JSA}
\end{equation}
where $\Gamma (\omega_s+\omega_i)$ is  the \emph{Pump Envelope Function} (PEF) and $\Phi(\omega_s,\omega_i)$ is the \emph{Phase Matching Function} (PMF), given by:
 
 \begin{equation}
     \Phi(\omega_s,\omega_i) =\frac{\sin(|\Delta \vec{k}|L/2)}{|\Delta \vec{k}| L/2}\label{Eq:PMF}
 \end{equation}
 
\noindent where $\Delta \vec{k}$ is the wave-vector mismatch, calculated as $\Delta \vec{k}=\vec{k_p}-\vec{k_s}-\vec{k_i}$, where $\vec{k_j}$ are the wave vectors of pump, signal and idler photons respectively. The wave-vectors are depicted in Figure \ref{fig1_PA}(b): each of them is along the photon propagation direction, and has magnitude $|\vec{k_j}|= 2\pi n_j/\lambda_j$, where $\lambda_j$ is the wavelength and $n_j$ the refractive index at the corresponding frequency.

It is worthwhile to note the similarity between SPDC and OPA, which are fundamentally driven by same physical mechanisms. OPA can be very robustly described using the classical treatment of electromagnetic waves and nonlinear polarization. A very important descriptor of the OPA mechanism is the efficiency of the energy flow from the pump to the signal/idler fields, which can also be written as a function of the wave vector mismatch~\cite{manzoni2016design}. Notably, the efficiency of the nonlineat OPA process acquires a functional form, which is similar to Eq.~\ref{Eq:PMF}, which further highlights the common origin of SPDC and OPA and the equivalence of the quantum and classical treatments in nonlinear optics. 

%

Importantly, irrespective of the process, for each $\{\omega_s,\omega_i\}$ pair (and the corresponding pump $\omega_p=\omega_s+\omega_i$), the PMF strongly depends on $\Delta \vec{k}$, the \emph{wave-vector mismatch}, and on the crystal length $L$. As anticipated, the PMF determines the frequency range of the emitted signal/idler photons, their polarization, their emission direction, and, in close relation with the subject of this paper, their degree of spectral entanglement. For this reason, we will now discuss the different phase matching configurations.

\subsection{Phase Matching condition}

\begin{figure*}
    \includegraphics[width=.8\textwidth]{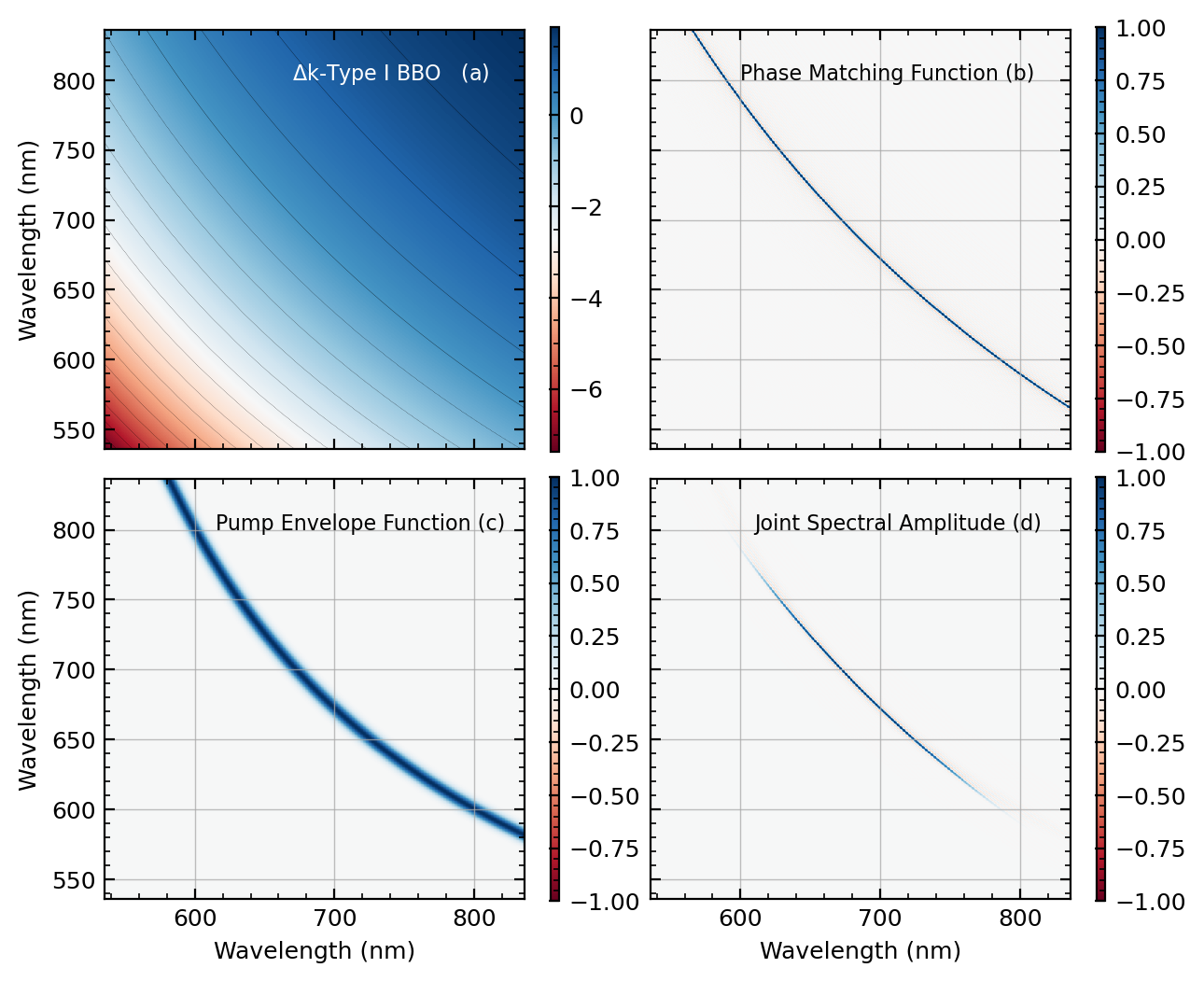} 
    \caption{Simulations for a 1-mm-thick BBO with $\alpha =0$° (collinear configuration) and $\theta =34.2$° with a 343\,nm pump beam and in Type I configuration (ooe). PM condition (a), PMF (b), PEF (c) and JSA (d).}
    \label{fig2_Collinear}
\end{figure*}

Equation \ref{Eq:Eff} states that the parametric downconversion is optimal when $\Delta \vec{k}=\vec{k_p}-\vec{k_s}-\vec{k_i}=0$, a condition which is termed \emph{Phase Matching} (PM). For a given set of pump, signal, and idler frequencies, PM is obtained by adjusting the propagation directions $\alpha$ and $\beta$ of signal and idler photons with respect to the pump, and by tuning their refractive index. We consider uniaxial birefringent crystals, which are characterized by two orthogonal polarization directions, the ordinary ($o$) and extraordinary ($e$): when a photon is polarized along the $o$ direction, it experiences a refractive index $n_o$, while for the polarization along the $e$ direction the refractive index is $n_e(\theta)$, where $\theta$ is the angle between the propagation direction $\vec{k_j}$ and the optical axis of the crystal. 

Figure \ref{fig1_PA}(b) shows the optical axis of the crystal and the angles it forms with the pump and the signal/idler photons, which are respectively $\theta$, $\theta+\beta$ and $\theta-\alpha$. By acting on the angle $\theta$ between the pump wave vector and the optical axis and on the polarization of the interacting photons, which determines the corresponding refractive indexes, it is possible to tune $\Delta \vec{k}$ to optimize the process efficiency and obtain the so-called \emph{Angle-Tuning Phase Matching}. The polarization state of the photons can be grouped in the following configurations (see table \ref{TAB_Type}):

\begin{itemize}
\item Type 0: the three interacting photons have equal polarization states, either ordinary (ooo) or extraordinary (eee).
\item Type I: the signal/idler photons propagate with the same polarization, either ordinary or extraordinary, but perpendicular with respect to the pump. The resulting configurations are therefore ooe or eeo, respectively.
\item Type II: the signal/idler photons are cross-polarized, and the possible configurations are eoe, oee, eoo, oeo.
\end{itemize}
\begin{table}
\centering
\caption{Summary of polarization configurations in second-order processes.}
\begin{tabular}{|c|c|c|}
\hline
& $\omega_s, \omega_i, \omega_p$(e)
& $\omega_s, \omega_i, \omega_p$(o) \\
\hline \hline
Type 0& \emph{e e e} & \emph{o o o}\\
\hline
Type I& \emph{o o e} & \emph{e e o}\\
\hline
Type II& \emph{e o e} & \emph{e o o}\\
& \emph{o e e} & \emph{o e o}\\
\hline
\end{tabular}
\label{TAB_Type}
\end{table}

\begin{figure*}
    \includegraphics[width=.7\linewidth]{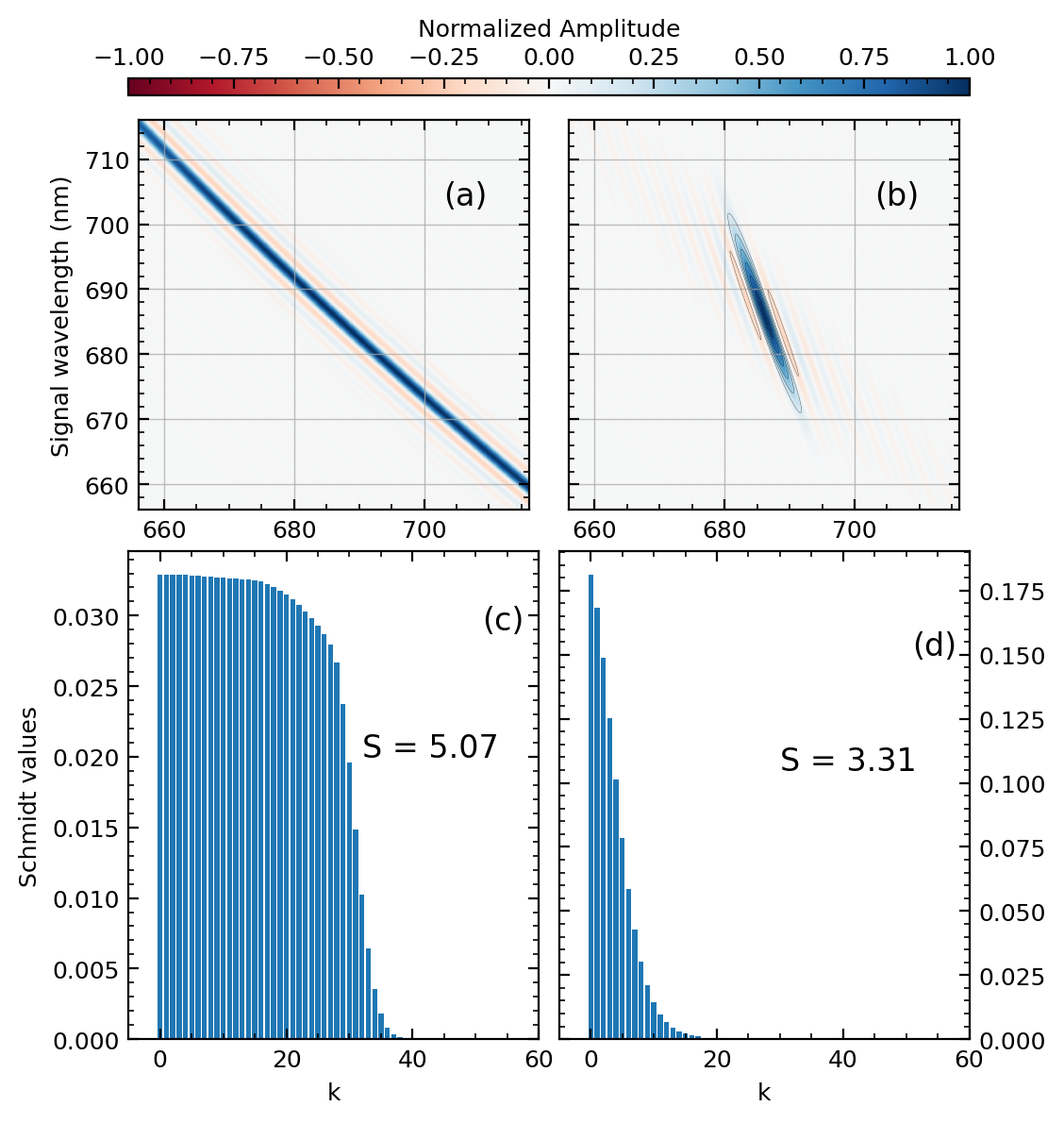} 
    \caption{JSA (a), (b) and Schmidt values (c), (d) respectively for Type I and Type II interactions in BBO for non collinear geometry ($\alpha=2.5$° and $\theta$ of $34.9$ and $53.8$ respectively), corresponding to the experimental condition we want to exploit. The von Neuman entropy, $S$ for each of the cases is given in (c) and (d).}
    \label{fig3_Entropy}
\end{figure*}

\noindent Typically, the SPDC is phase-matched only for one of these configurations; hence the phase-matching angle and cut of the nonlinear crystal univocally determine the polarization states of the spontaneously emitted photons. 
Birefringent phase-matching can be obtained only with Type I and Type II interaction schemes. In addition, for a specific birefringent crystal only a fraction of  the configurations listed in Table \ref{TAB_Type} can be phase-matched. In particular, for negative (positive) uniaxial crystals PM can only be obtained if $\omega_p$ is extraordinary (ordinary). For instance, $\beta$-barium borate (BBO), which is one of the most widespread nonlinear media and will be the medium used in the simulations and experiments of this paper, is a negative uniaxial crystal; hence, PM requires extraordinary polarization for $\omega_p$.

For anisotropic nonlinear crystals, the polarization of the interacting fields not only affects the phase mismatch, but also the nonlinearity of the medium, by selecting a specific element of the nonlinear susceptibility $\chi^{(2)}$ tensor \cite{manzoni2016design}. In some nonlinear crystals, the highest nonlinearity occurs when the interacting beams have the same polarizations; this corresponds to the Type 0 configuration, which in turn does not enable birefringent phase-matching. This is the case of Stoichiometric Lithium Tantalate (SLT) or Lithium Niobate (LiNbO$_3$, LN). In this configuration, one may adopt the so-called \emph{Quasi-Phase Matching} (QPM) configuration, which introduces a spatial modulation in the nonlinear medium by periodically inverting the sign of the nonlinear coefficient $\chi^{(2)}$, \textit{i.e.}, in the so-called periodically poled structures\cite{HUM2007180}. QPM enables the use of nonlinear crystals with large $\chi^{(2)}$ (10-20 pm/V) - one order of magnitude higher than the standard birefringent nonlinear media like BBO  - for which birefringent PM cannot be achieved. For the case of QPM the wave vector mismatch reads:
\begin{equation} \label{EQ_MomentumPP}
\Delta\vec{k}=\vec{k_p}-\vec{k_s}-\vec{k_i}-\vec{K_g}
\end{equation} 
where $|\vec{K_g}|=2\pi/\Lambda$ is the grating wave vector and $\Lambda$ is the spatial modulation period of the nonlinear coefficient. From this equation, illustrated in Fig.~\ref{fig1_PA}(c), one directly calculates the poling period $\Lambda$ that fulfils phase-matching ($\Delta\vec{k}=0$).

In the next sections, we will estimate the phase-matching function for various crystal configurations and investigate the optimal configuration for broadband spectral entanglement. 

\subsection{Joint Spectral Amplitude}

The measurable characteristic of a biphoton state, as written in Eq.~\ref{Eq:biphoton}, is the biphoton field amplitude, $\mathcal{F} (\omega_i,\omega_s)$, also referred to as the JSA.  For the sake of convenience, in the following, we will refer to the wavelengths of the SPDC photons instead of their frequencies. The JSA corresponds to the probability of the signal photon to be at a given wavelength $\lambda_s$ at the exact moment when an idler photon is at $\lambda_i$. This two-dimensional correlation function between the signal and idler wavelengths quantifies the degree of entanglement within the biphoton state. The JSA is expressed as the product of the PEF, $\Gamma (\lambda_s,\lambda_i)$ and the PMF, $\Phi(\lambda_s,\lambda_i)$. The latter is calculated by first estimating the phase-matching condition Eq.\ref{EQ_MomentumPP} for a given crystal and {$\lambda_p,\lambda_s,\lambda_i$} triplet, and then using Eq.~\ref{Eq:PMF}.   

We consider the SPDC process driven by a pump pulse at a wavelength $\lambda_p =$ 343\,nm and a bandwidth of 3\,nm. This corresponds to the typical bandwidth of the third harmonic of the output of a Ytterbium-based femtosecond laser oscillator. This yields a PEF, $\Gamma$, shown in Fig.~\ref{fig2_Collinear}(c), which accounts for the signal/idler pairs fulfilling the energy conservation as in Eq.~\ref{Eq: SPDC1}. We first consider a collinear configuration ($\alpha = 0$° in Fig.~\ref{fig1_PA}(b)) where signal and idler photons are emitted in the same direction of the pump pulse. We also consider a 1-mm-thick BBO cut for Type I (ooe) interaction at degeneracy ($\overline\lambda_s = \overline\lambda_i=686$nm). This corresponds to an optimum phase-matching angle $\theta=34.2$°; signal and idler photons will have parallel ordinary polarization. 

The wave vector mismatch $\Delta \vec{k}=\vec{k_p}-\vec{k_s}-\vec{k_i}$ calculated for this crystal and pump configurations at each signal and idler wavelengths is shown in Fig.~\ref{fig2_Collinear}(a). Given the collinear interaction, $\Delta \vec{k}$ is a scalar ($\Delta k$) with positive and negative values. Note that momentum conservation is obtained also for the signal-idler pairs, which are away from the degeneracy condition. The PMF $\Phi(\lambda_s,\lambda_i)=\eta(\lambda_s,\lambda_i)$, which depends on $\Delta k$ and the crystal thickness (see Eq.~\ref{Eq:PMF}) is shown in Fig.~\ref{fig2_Collinear}(b). The JSA $\mathcal{F}(\omega_1,\omega_2)$ is obtained as the product between the PMF and the PEF (shown in Fig.~\ref{fig2_Collinear}(c)) and is shown in Fig.~\ref{fig2_Collinear}(d).

\subsection{Entanglement Entropy}

The JSA in Fig.~\ref{fig2_Collinear}(d) measures the degree of entanglement carried by the down-converted photons. JSA can be factorized as a tensorial product of two inner product spaces, which define the eigenspaces of each of the photons. This can be done numerically via a singular value decomposition (SVD), also known as Schmidt decomposition, as:

\begin{equation}
    \mathcal{F}(\omega_s,\omega_i) = \sum_n r_n f_n(\omega_s)g_n(\omega_i)
    \label{Eq_SVD}
\end{equation}

\noindent Considering the eigenvalues, or the Schmidt coefficients, $r_n$, normalized to unity, the entanglement, or Von Neumann entropy is estimated as

\begin{equation}
    S = -\sum_n r_n \ln r_n
\end{equation}

For example, in Fig.~\ref{fig3_Entropy}(a) and (b), we show the JSA estimated for Type I and Type II BBO respectively, considering the best phase-matching ($\theta$ of $34.9$ and $53.8$ respectively) in the non-collinear geometry for $\alpha=2.5$°. We restrict the detection region, for the sake of the discussion, to a bandwidth of 30\,nm. Experimentally, this corresponds to the introduction of spectral bandpass filters with a \emph{top-hat} transmission spectrum. 
The JSA maps can be decomposed using the singular value decomposition algorithm to estimate the associated Schmidt components, whose associated singular values ($r_n$ in Eq.~\ref{Eq_SVD}) are reported in Fig.~\ref{fig3_Entropy}(c) and (d) for Type I and Type II cases respectively. 

To be more precise, entropy is calculated using a numerical singular value decomposition algorithm applied to the JSA maps. The resulting values (Singular Values, $SV$) in the diagonal matrix are normalized ($SV_N$) such that the sum of their squares is equal to 1 (normalization of the Eigenvalues): $SV_N=SV/\sqrt{\sum_n(SV^2)}$ so that $\sum_n(r_n)=\sum_n[(SV_N)^2]=1$. Entropy is then calculated as $S=-\sum_n\{(SV_N)^2 log_2[(SV_N)^2]\}$. The Schmidt Coefficients plotted in Fig~\ref{fig3_Entropy}(c) and (d) are the Normalized Singular Values ($SV_N$) to the power of 2 (i.e. the Eigenvalues). Another figure of merit of the separatability of the biphoton state is the Schmidt Number, which is calculated as $K=1/\sum_n[(SV_N)^4]$ ~\cite{zielnicki2018joint,Eberly2006,Law2000}. 

For Type I, we observe the spectral correlation between the photon-pair over the entire bandwidth, as evident in the diagonal feature over the 2D JSA map shown in Fig.~\ref{fig3_Entropy}(a). This is also substantiated by a large number of Schmidt components, quantified by the entanglement entropy, which is estimated to be 35.8 for this case. Type II presents a completely different scenario, where the two photons are generated only close to the degeneracy point and with weaker spectral correlation than in the Type I case. Accordingly, a much lower entanglement entropy of 10.17 is estimated in this condition. These two examples clearly demonstrate that the specific choice of the PM condition can result in very distinct entanglement entropies, with Type I favoring maximal entanglement and Type II generating photon states with minimal spectral correlation, that result in the so-called heralded single photons~\cite{mosley2008heralded}. Fundamentally, this implies that a larger phase matching bandwidth results in greater entanglement entropy. 

\begin{figure}
    \centering
    \includegraphics[width=8 cm]{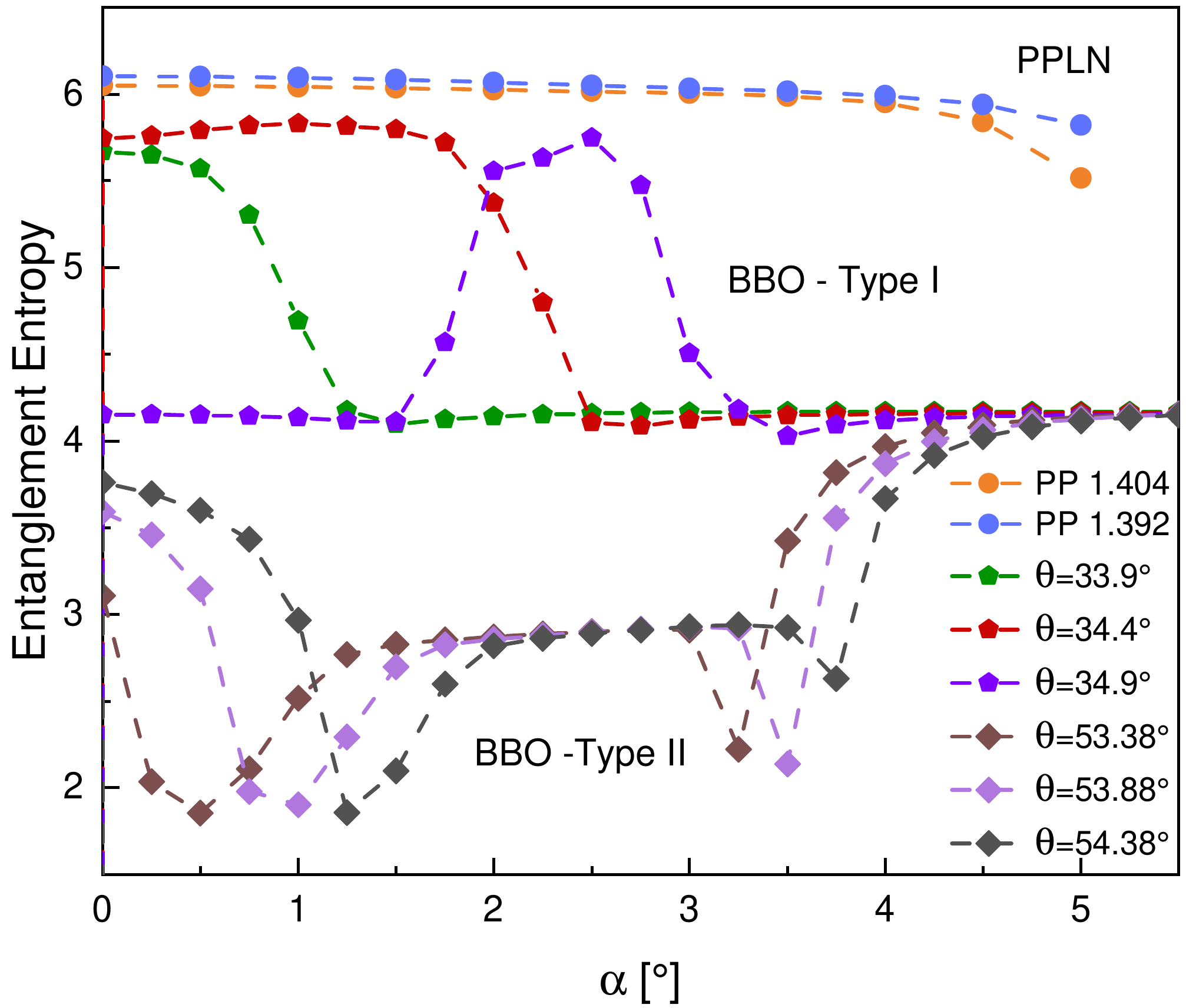}
    \caption{Entropy calculated on SPDC maps with gradual variation of angle $\alpha$ from 0$^o$ to 5$^o$ with 0.25$^o$ steps. Different configurations of PM are considered: collinear geometry in PPLN with 2 different poling periods (PP), Non-collinear Type I in BBO crystals and Non-Collinear Type II in BBO crystals with different $\theta$ angles.}
    \label{fig4_AngleDep}
\end{figure}

In addition to the polarization configurations, we can also employ two phase-matching angles to optimize the entropy. These correspond to the angle between the crystal optical axis and the incident pump wave vector ($\theta$ in Fig.~\ref{fig1_PA}) and the angle between the pump and signal wave vectors ($\alpha$). We explore this dependence in Fig.~\ref{fig4_AngleDep}, where the value of the entanglement entropy estimated through the SVD analysis of the theoretical JSA maps is plotted for various values of $\theta$ and $\alpha$ and for both Type I and Type II configuration. Type I condition results in higher entropy values at all angles, but clear peak values can be observed at specific crystal cuts quantified by $\theta$. For a given value of $\theta$, it can be seen that the maximum entropy value is observed for spatial photon modes emitted over a relatively narrow range of angles. While on one hand this suggests that the collection angle can be used as an effective experimental knob to tune the entropy for a given crystal, it also highlights the importance of precision in the collection to maximize the entropy. 

We note that the value of the entanglement entropy appears to plateau to the same value for large emission angles and for both Type-I and Type-II cases. The entropy values are a direct outcome of numerical analysis of the two-dimensional JSA maps, without any analytical simplification. Accordingly, the estimated values are an interplay between the PMF and PEF maps. Non-trivial trends can be seen in Fig.~\ref{fig4_AngleDep} for those conditions where the maximum of the PMF is still within the range of the PEF (see the SI for more in-depth analysis of the JSA maps). At larger emission angles however, only the tails of the PMF still remain within the considered range, in which case the JSA function is primarily determined by the shape of the PEF, thus resulting in the same value for the entropy irrespective of the crystal geometry. However, we do not expect efficient generation of the SPDC photons in such cases and we also note that the brightness of the photons is not explicitly considered in the presented formalism.

Interestingly for the Type II condition, the entropy value is lowered over a narrow range of $\alpha$ angles. Over a relatively wide range of emission angles the entropy plateaus to a constant , which is fundamentally determined by the spectral bandwidth of the pump and filter before the detectors (determined by the restriction of the explored wavelength region in the simulation). The angular dependence of the entropy is strongly reduced in the case of Type 0 PM achieved in periodically poled LN crystals. The entropy values for two poling periods are shown in Fig.~\ref{fig4_AngleDep} and we observe substantially improved entropy in both cases. This suggests that periodic poling is an ideal way to achieve maximal spectral entanglement with respect to using birefringent phase matching. A periodically poled crystal at degeneracy presents a scenario in which the group velocities of signal and idler photons are matched and this maximizes the phase matching bandwidth. Accordingly, the entanglement entropy is also maximized. Note that the calculations presented above do not estimate the brightness of the SPDC sources, which is an additional factor to be considered. 

The theoretical formalism discussed in this section provides experimental handles to tune the energy and entanglement entropy of spectrally entangled biphoton states. The challenge now is to measure their quantum characteristics, particularly those that can be effective optical probes of nonlinear processes in materials. In the next section, we provide our perspective on nonlinear spectroscopy with quantum entangled photons and propose and demonstrate an experimental scheme for characterizing the JSA, which can be applied to spectroscopy of molecular materials. 

\begin{figure*}
    \centering
    \includegraphics[width=0.9\linewidth]{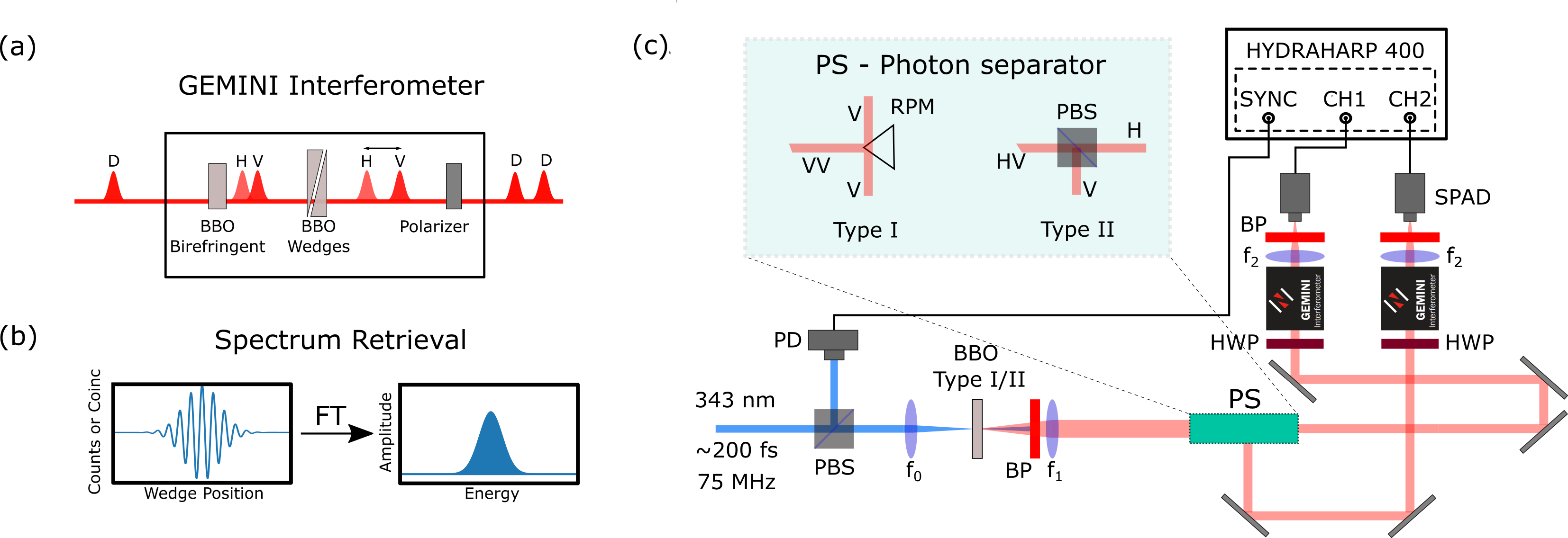}
    \caption{Schematic representation of (a) GEMINI interferometer, (b) 1D-Fourier analysis, and (c) Experimental setup: The pump pulses (343\,nm, $<$200\,fs) are the third harmonic of a mode-locked Yb:KGW oscillator running at 75\,MHz (Pharos, Light Conversion) and are generated using a Free-Standing Harmonic Generator (HIRO, Light-Conversion). The output polarization is cleaned using a polarizing beam splitter (PBS); the residual light is sent to a photodiode, PD (TDA 200, PicoQuant) and it is used as a trigger for the time-correlated single-photon counter (HydraHarp 400, PicoQuant). For the Type II BBO crystal,  a 15\,cm fused silica lens ($f_0$) focuses the pulse, the  UV pump is removed with a filter (BP) centered at 700\,nm with 40\,nm band and the SPDC photons are then collimated with a 5\,cm lens ($f_1$). The two photons with orthogonal polarization are separated with a second PBS. For the Type I BBO crystal the focusing lens ($f_0$) is a 30 cm fused silica and a 10 cm lens is used for collection ($f_1$), the two photons are separated using a right angle prism mirror. In both cases, after separation, each photon is sent to a TWINS interferometer (GEMINI, NIREOS). Prior to the interferometer, a half-waveplate (HWP) is used to rotate the polarization to 45$^\circ$ to minimize the losses. Each photon is focused by a 7.5\,cm lens ($f_2$) into a SPAD detector (MPD) with a 50\,{\textmu}m active area. The eventual background is removed with the use of another bandpass filter (centered at 700\,nm, 40\,nm band) before each detector.}
    \label{SetupSketch}
\end{figure*}

We note here that the phase-matching conditions considered above involve non-collinear configurations, while Eq.~\ref{Eq:JSA} is derived assuming collinear condition (see SI). However, we do not expect that the definition of the joint spectral amplitude as the product of the phase matching function and the pump envelope function will change moving from the 1D collinear to
3D non-collinear scenario. In the latter case, the JSA function will intrinsically acquire a spatial dependence, mirroring the dependence of the phase-matching function. All our calculations assumed a single spatial mode for the SPDC photons and thus the spatial dependence of the JSA can be ignored.

\section{Entangled photon spectroscopy - measurement principles}\label{Sec:Experiment}

One of the primary goals of nonlinear spectroscopy is to obtain insights into the nature and dynamics of many-body interactions in materials. Typically, such interactions are mirrored in the nonlinear susceptibility of a photo-excited sample, which generates a nonlinear polarization that subsequently emits coherent radiation. Measurement of the spectral amplitude and phase of this coherent wave provides all the information about the system-bath interactions. For example, some of us have previously discussed nonlinear spectral lineshapes in the presence of exciton-exciton scattering in a two-dimensional semiconductor, see Ref.~\citenum{srimath2020stochastic}. 

As noted in the introduction, the goal of entangled photon spectroscopy is to extract the same information, but at single photo-excitation levels, thus for the lowest possible sample perturbation. In our perspective, this can be achieved by measuring the change induced in the JSA of the biphoton state ($\mathcal{F}(\omega_i,\omega_s)$ in Eq.~\ref{Eq:biphoton}) due to its interaction with the sample. The specific configuration of the interaction can give us access to the first and second-order responses. In the simple scenario of first-order scattering, when one of the photons is transmitted through the sample of interest, the JSA function of the biphoton field is effectively multiplied by a linear response function. This is equivalent to performing linear absorption spectroscopy with classical light. However, given the strong quantum correlations between the photons, we envisage superior measurement sensitivity, that may enable the detection of chromophores at extreme dilutions. This may also be expanded to transient absorption schemes with the biphoton state replacing the classical probe pulse.   
Alternatively, when both the photons are transmitted through the sample, Bittner et al~\cite{bittner2020probing} have theoretically demonstrated that the consequent second-order scattering can be modeled as a product of the JSA function of the incident state and a two-photon scattering amplitude, $\mathcal{S}^{(2)}$. Following Ref.~\citenum{li2019photon}, the transmitted state can be written as:

\begin{widetext}

\begin{equation}
    |\psi_{out}\rangle = \frac{2\pi A}{i\hbar}\int \int d\omega_id\omega_s \mathcal{F}_{out}(\omega_i,\omega_s) \hat{a_i}^{\dagger}(\omega_i) \hat{a_s}^{\dagger}(\omega_s)|0\rangle,
    \label{Eq:output}
\end{equation}
where
\begin{equation}
\mathcal{F}_{out}(\omega_i,\omega_s)=\int \int d\omega'_id\omega'_s \mathcal{S}^{(2)}(\omega_i,\omega_s;\omega'_i,\omega'_s)\mathcal{F}_{in}(\omega'_i,\omega'_s)
     \label{Eq:output1}
\end{equation}
with
\begin{equation}
    \mathcal{S}^{(2)} (\omega_i,\omega_s; \omega'_i,\omega'_s) = \mathcal{S}^{(1)} (\omega_i,\omega'_s)\mathcal{S}^{(1)}(\omega'_i,\omega_s) e^{-g_{is}},  
    \label{Eq:two-scat}
\end{equation}
\end{widetext}

\begin{figure*}
    \centering
    \includegraphics[width=0.75\linewidth]{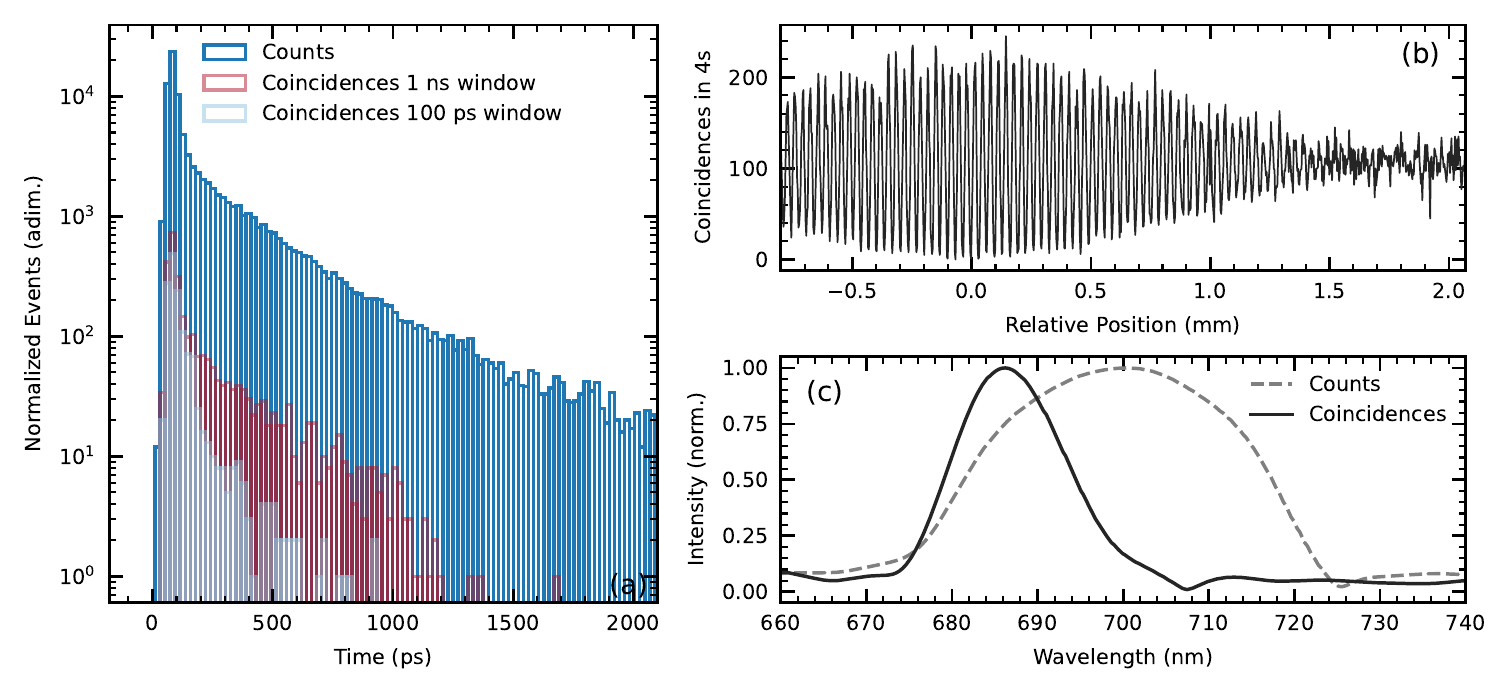}
        \caption{(a) Single count distribution recorded in T3 mode (blue) and coincidences count computed with a time window of 1\,ns (red) and 100\,ps (grey). (b) Interferogram of coincidences retrieved by scanning the TWINS on one arm. (c) Comparison of FT retrieved spectrum of the SPDC photons: dashed line is from the interferogram of total photon counts while the continuous line is from the interferogram of coincidence counts.}
    \label{fig_counts}
\end{figure*}
where $\mathcal{S}^{{(1)}}$ represents the single photon scattering and $g_{is}$ is a cumulant function that accounts for cross correlation between the photon modes. Such cross-correlations cannot be induced through light--matter interactions where the material systems has no intrinsic nonlinearity. Systems with many-body interactions, such as two-dimensional materials with exciton--exciton scattering, present a way to induced such correlations. Consequently, measurement of the JSA function of the transmitted biphoton state, $\mathcal{F}_{out}(\omega_i,\omega_s)$ which is determined by many-body interactions in the material, and comparison with the input JSA, $\mathcal{F}_{in}(\omega_i,\omega_s)$, can directly provide us access to the many-body interaction parameter, much like the two-quantum coherent excitation spectroscopy~\cite{Thouin2018}, but at extremely low photon flux. However, the prerequisite is an experimental strategy that can reliably measure $\mathcal{F}(\omega_i,\omega_s)$, which we will discuss in the following. 

Simply put, one has to estimate the probability of detecting the signal photon at a specific frequency ($\omega_s$) given that the idler photon is at a complementary frequency ($\omega_i$). This can be done in a rather straightforward way by selecting the frequency of each of the photons using a scanning monochromator before the detection stage. As noted by Zielnicki et al.~\cite{zielnicki2018joint}, the intrinsically low efficiency of commercial monochromators and the scanning methodology make the measurements inefficient. An alternative methodology based on Fourier transform (FT) spectroscopy was proposed~\cite{zielnicki2018joint, maclean2019reconstructing, maclean2018ultrafast}, in which each of the photons is transmitted through a scanning Michelson interferometer, and a time-domain measurement of joint spectra is performed.

Here we perform FT spectroscopy using a common-path birefringent interferometer, the Translating-Wedge based Identical pulses eNcoding System (TWINS)~\cite{Brida:12}, schematically shown in Fig.~\ref{SetupSketch}(a). TWINS consists of two birefringent wedges, with optical axes perpendicular to each other, placed between two polarizers that have their transmission axis set at 45$^o$ with respect to the optical axes. In the classical perspective, the first polarizer selects the diagonal polarization state of the incoming electromagnetic wave, which is composed of equal components of horizontal and vertical polarization.  The wedge pair imparts a tunable delay between these two polarization components. Effectively, one is making two identical copies of the incident field and imposing a phase delay between them, much like in a Michelson interferometer. These components are projected to a common polarization by the second polarizer and interfere, so that the output intensity accordingly oscillates as the delay is scanned, through the position of the wedge. 

The FT of the recorded interferogram gives the spectrum of the incident electromagnetic wave, as indicated in Fig.~\ref{SetupSketch}(b). This system has been employed to measure the steady-state and transient spectral response of various light sources and it is made commercially available as the \textit{GEMINI} interferometer from NIREOS s.r.l., described elsewhere.~\cite{Oriana:16} With respect to a standard Michelson interferometer, TWINS offers the advantages of compactness and exceptional delay precision, stability and reproducibility. In the following, we demonstrate that FT spectroscopy based on TWINS can be effectively employed to estimate the JSA of the spectrally entangled state generated in a Type I and Type II BBO crystal.

\subsection{Experimental methodology}\label{subsec:exp_met}

\begin{figure*}
    \centering
    \includegraphics[width=0.75\linewidth]{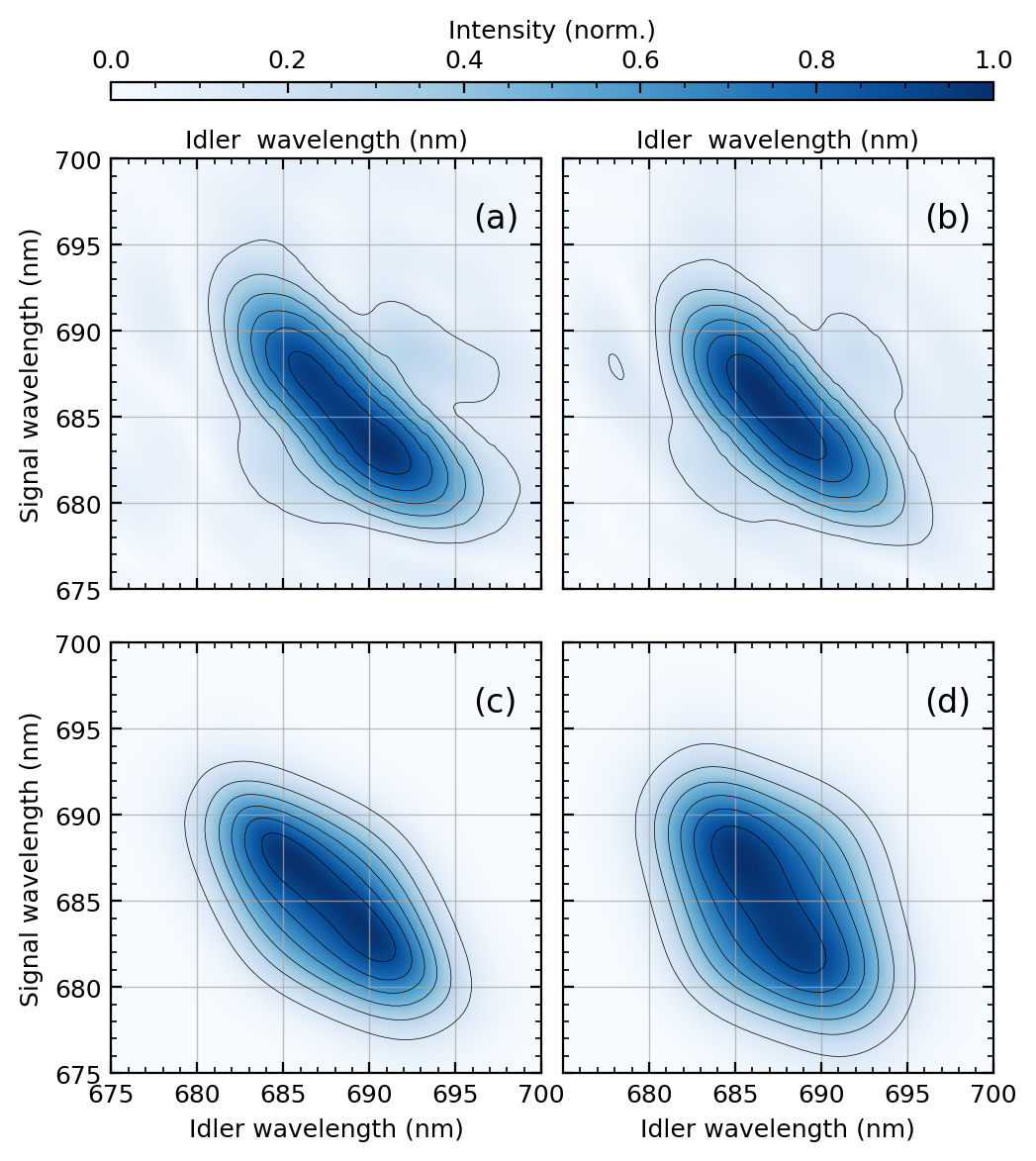} 
    \caption{Experimental JSI measured using two TWINS interferometers for (a) a Type I and (b) Type II BBO crystal. Simulated JSI for Type I (c) and Type II (d) BBO crystal, obtained from a linear sum of calculated JSI with $34.6^o <\theta < 35.2^o$ and $53.4$° $<\theta < 54.2$° respectively. The collection angle of the emission is kept in the range $2.25$° $<\alpha < 2.75$°.}
    \label{figJSI}
\end{figure*}

Figure ~\ref{SetupSketch} shows the experimental scheme employed to characterize the JSA of SPDC photons generated in a Type II or a Type I BBO crystal. For the sake of clarity, we will first introduce the methodology used to detect the photons before discussing the use of the TWINS. Accordingly, we first consider the scenario without the TWINS in the scheme.

The SPDC photon pairs generated in the Type II BBO crystal, which are orthogonally polarized, are separated by a polarizing beam splitter (PBS) and then sent into the single-photon avalanche diodes (SPADs) after the bandpass filters. For the Type I BBO crystal, since the photons have the same polarization, the separation happens spatially using a right angle prism mirror placed at 10 cm from the collection lens.
The photon detection events are recorded by a photon counter in which the events are time-tagged with respect to the sync events from the reference photodiode (PD) at the 75-MHz repetition rate of the laser oscillator. The time-tagged events from both detectors are logged and analyzed on the fly for coincidence events by a homemade LABVIEW routine. Two logged events are considered coincident if they belong to the same sync and are recorded within a pre-selected time window. This coincidence time window is determined by considering the response time and the jitter of the detectors and the photon counting unit, with the goal to minimize accidental coincidence events from background emission from the BBO crystals (see discussion below).

Generation of SPDC photons in BBO occurs only when the pump field is present in the crystal and accordingly photons can only be observed close to the zero-time. However, the temporal histogram of events from one of the detectors, shown in Fig.~\ref{fig_counts}(a) (blue), has a long living tail extending into the nanosecond range. Note that the dynamics here are not limited by the resolution of the detection system, which is estimated to be approximately 80\,ps. We attribute the additional, delayed detection events to the photoluminescence (PL) from color centers in BBO, whose spectral distribution, unfortunately, overlaps with that of the SPDC photons. Such emission processes are typically uncorrelated and thus will not result in coincidence detection unless the detection window is long enough to accidentally capture two simultaneous PL events on both detectors. 
The appropriate detection window can be assessed by analyzing the temporal distribution of coincidence events, as can be seen from Fig.~\ref{fig_counts}(a), where we show coincidence histograms taken with 1\,ns and 100\,ps  time windows. We observe a substantial tail in the case of 1\,ns window, which can be attributed to accidental coincidences and thus not representative of SPDC photons. With a stricter 100\,ps window, we remove most accidental events and observe coincidences predominantly close to the zero-time, as expected from SPDC photons. We use this time window to define coincidence detection with TWINS in place for the JSA estimation.  Further details on the background and accidental coincidences are reported in the Supplementary Information \ref{sec:SI2}.

Before the JSA measurement, we will briefly discuss the incorporation of the TWINS into the experiment. As explained earlier, TWINS can be used in lieu of conventional Michelson interferometers to measure the spectrum of an incident electromagnetic field in the time domain by FT spectroscopy. Extending such a measurement basis to the level of single photons, TWINS is introduced in the path of one of the photons and the coincidence rate is recorded as a function of the position of the wedge-pair. The resulting interferogram of coincidences is shown in Fig.~\ref{fig_counts}(b). A discrete FT (DFT) algorithm is then applied to retrieve the spectrum of the SPDC photons, shown in Fig.~\ref{fig_counts}(c). The peak of the spectrum is around 686\,nm, as expected for photons generated from degenerate SPDC of pump photon at 343\,nm. We highlight that the measurement of the spectrum using the total count rate from the detector yields a completely different result, shown in the dashed line in Fig~\ref{fig_counts}(c). This is essentially the spectrum of the broad background emission from the defects in the BBO, filtered by the 40\,nm bandpass filter. The algorithm we used to estimate the coincidences intrinsically eliminates the uncorrelated photons and effectively detects only the entangled SPDC photons. 

Finally, we introduce TWINS on the paths of both  photons and perform coincidence measurements by scanning the position of the wedge-pair in both interferometers. In this case, the coincidence rate as a function of the delays introduced by the two TWINS interferometers (which we call $\tau_s$ and $\tau_i$) can be written as:

\begin{equation}
\begin{aligned}
    C(\tau_s,\tau_i) 
    = \int \int d\omega_id\omega_s |\mathcal{F}(\omega_i,\omega_s)|^2 \times\\
    \times(1+cos(\omega_s\tau_s))(1+cos(\omega_i\tau_i))
    \label{Eq:output2}
    \end{aligned}
\end{equation}
where $\mathcal{|F}(\omega_i,\omega_s)|^2$, or the modulus square of the JSA, is the so-called Joint Spectral Intensity (JSI). This measurement results in a two-dimensional interferogram, which can be Fourier transformed using a 2D DFT algorithm to estimate the JSI of the SPDC photons. We note that the amplitude of the JSA can be obtained by simply taking the square root of the JSI, while determination of the phase requires more sophisticated experimental approaches. We measured the JSI for SPDC photons emitted by Type I and Type II BBO crystals with designs that should theoretically result in the JSI maps shown in Fig.~\ref{fig3_Entropy} (a) and (b) respectively. The experimentally determined JSI maps are shown in Fig.~\ref{figJSI} (a) and (b). The peak of the spectral correlation is close to the degeneracy point, at around $\lambda =$ 686\,nm, as expected for these particular crystal designs. The diagonal linewidth of the spectrum is mostly limited by the 40\,nm bandpass filters placed before the detectors (see Fig.~\ref{SetupSketch}). Despite this consideration, it is evident that the experimental measurements do not match the simulated spectra. In fact, entanglement entropy values are estimated from the experimental JSI maps as 0.78
(Schmidt number = 1.36) 
for Type I and 0.7 for Type II (Schmidt number = 1.3). This is particularly evident in the substantially larger anti-diagonal width in both spectra, which is eventually detrimental to the entanglement entropy. We note that this is not due to the spectral bandwidth of the pump pulses, which is approximately 4.5\,nm and thus not very different from the value used for the calculations.  

To rationalize this discrepancy, we recall the strong dependence of the entropy values on the angle of incidence ($\theta$) and the angle of emitted photons ($\alpha$), as shown in Fig.~\ref{fig4_AngleDep}. For the calculations shown in Fig.~\ref{fig3_Entropy}, we considered very specific values for both these parameters ($\alpha = 2.5^o$ and corresponding phase matching angle $\theta$). However, experimentally, we are rather sampling a range of values. Firstly, we are focusing the pump with a long focal length lens, which will lead to a range of incidence angles. Secondly, we are not restricting the collection to the photons emitted at $\alpha = 2.5^o$. We are rather integrating over a range of emission angles. Thus, a direct comparison between the experimental data in Fig.~\ref{figJSI} and calculations in Fig.~\ref{fig3_Entropy} is inaccurate. For a more accurate comparison, we calculated the JSI spectra for $34.6^o <\theta < 35.2^o$ for Type I phase matching and $53.3$° $<\theta < 54.3$° for Type II phase matching, with the collection angle of the emission in the range $2.25$° $<\alpha < 2.75$°. These values represent the experimentally sampled range of parameters, where the different $\theta$ span is to take into account the different focal lengths. By performing a linear sum of these set of JSI spectra, we obtained the JSI spectra shown in Figs.~\ref{figJSI}(c) and (d). Note that the summed JSI spectra are also multiplied with the spectral response of the bandpass filters. The spectra qualitatively reproduce the experimentally observed lineshapes, particularly with larger anti-diagonal linewidth although deviations still persists. Entropy for the simulated maps is equal to 0.6 (Schmidt number = 1.28) for Type I and 0.33 for Type II (Schmidt number = 1.12). There is a reasonably good match for Type I between the experimental entropy and simulated one, while for Type II the difference is larger. The discrepancy in the calculations can be attributed to the lack of SPDC efficiency's angular dependence in our theoretical formalism, which is inevitably present in the experiment.

\color{black}



 \section{Conclusions}

Spectroscopic methods that exploit the quantum nature of light may offer ways to achieve a significant improvement in the signal-to-noise ratio needed to observe coherent dynamics in condensed matter in the low-excitation-density limit. Here, we discussed one strategy based on entangled photons, which can be an effective counterpart to conventional nonlinear spectroscopy, in the quantum limit. Firstly, we presented an extensive review on the nonlinear optics of the SPDC process, which is used extensively to generate quantum entangled photons. Following this established framework, we developed design principles to generate spectrally entangled biphoton states with tunable energy and bandwidth. As a first step towards our experimental approach to quantum spectroscopy of materials, where the spectral correlations between the entangled photons are used as optical probes of many-body interactions, we presented an experimental method that can be used as a spectroscopic platform. The fundamental basis of our scheme is the estimation of JSI of the biphoton state using Fourier transform spectroscopy implemented via the TWINS interferometer. Our hope is that the quantum optical methodologies discussed here are used to perform nonlinear spectroscopy of materials at unprecedentedly low excitation densities and access novel quantum phenomena that are not accessible to classical optical probes. Application of the developed technique to a specific material problem is an ongoing effort and, due to additional complexity in the analysis, it will be presented in a separate publication.

\begin{acknowledgements}
ARSK acknowledges the start-up funds provided by Wake Forest University and funding from the Center for Functional Materials and the Office of Research and Sponsored Programs at WFU. We also thank Nick Bertone from Optoelectronic Components for supplying the photon-counting units in the initial stages of this project. The authors thank Prof. Carlos Silva and Prof. Eric Bittner for enlightening discussions and support and Prof. Daniele Faccio and Prof. Matteo Clerici for providing us the Type I BBO crystal and many useful discussions on the experimental scheme. 
\end{acknowledgements}

\section*{Author Contributions}

The simulations presented here have been performed by LM under the guidance of CM. The initial experiments were implemented by LM, CT, and ARSK. The final experimental design was developed and implemented by ERG, LM and LU. The experimental measurements were performed by LU, DOT, EK, and ERG. FP and AP supplied the TWINS interferometers. ARSK and GC conceived and supervised the project.  All the authors contributed to the development of the manuscript. 

\section*{Conflict of Interest}

FP, AP and GC disclose financial association with the company NIROES (www.nireos.com), which manufactures the TWINS interferometer used in this paper. 

\section*{Data Availability}

The data that support the findings of this study are available from the corresponding author upon reasonable request.
\section{References}
\bibliographystyle{apsrev4-2}
%

\clearpage
\newpage
\onecolumngrid
\textbf{\Large{Supplementary Information: Measurement principles for quantum spectroscopy of molecular materials with entangled photons}}
\vspace{1cm}%
\twocolumngrid

\makeatletter 
\renewcommand{\thefigure}{S\@arabic\c@figure}
\makeatother

\setcounter{section}{0}
\setcounter{figure}{0}
\section{\label{sec:SI}Supplementary Note 1: Parametric down-conversion in the interaction picture}
\noindent The temporal evolution of a state vector in the interaction picture can be written as:

\begin{equation}\label{tevolution}
    |\psi(t)\rangle = exp[{\frac{1}{i\hbar}}\int^t_{t_0}dt'\hat{H}_\mathrm{I}(t')]|\psi(t_0)\rangle ]
\end{equation}

\noindent where $|\psi(t)\rangle$ is the state vector at time $t$, $\hat{H}_\mathrm{I}(t)$ is the time-dependent interaction Hamiltonian, $\hbar$ is the reduced Planck constant, and $|\psi(t_0)\rangle=|\psi_0\rangle$ is the initial state vector at time $t_0$. For type II down-conversion, the interaction Hamiltonian can be expressed as:
\begin{equation}\label{Hinteraction0}
    \hat{H}_\mathrm{I}(t)=\int_{V}d^3r~\chi^{(2)}~\hat{E}_\mathrm{p}^{(+)}(r,t)~\hat{E}_\mathrm{o}^{(-)}~\hat{E}_\mathrm{e}^{(-)}+c.c.
\end{equation}
\noindent where $\chi^{(2)}$ is the second-order nonlinear susceptibility, $\hat{E}$ is the electric field, and the pedixes p, o and e stand for pump, ordinary and extraordinary, respectively. If we assume a collinear configuration, the electric fields along the propagation direction (z) can be expressed as:

\begin{equation}\label{Efield+}
    \hat{E}_\mathrm{j}^{(+)}(z,t)=\int d\omega_j~A(\omega_j)~\hat{a}_j(\omega_j)exp\{i[k_j(\omega_j)z-\omega_j t]\}
\end{equation}
\noindent where
\begin{equation}\label{vectorpotential}
    A(\omega_j)=i\sqrt{\hbar \omega_j}
\end{equation}
\noindent and:
\begin{equation}\label{Efield-}
    \hat{E}_\mathrm{j}^{(-)}(z,t)=[\hat{E}_\mathrm{j}^{(+)}(z,t)]^+.
\end{equation}
For an intense pump field (i.e. photon flux $>\SI{e13}{photons/s}$, corresponding to an average pump fluence  $>\SI{0.1}{\mu J/cm^2}$ at $\SI{1}{kHz}$ and photon energy $\SI{2.2}{eV}$), $E_\mathrm{p}$ can be safely treated as a coherent state:
\begin{equation}\label{Efield-2}
    \hat{E}_\mathrm{p}^{(+)}(r,t)\rightarrow E_\mathrm{p}(r,t)=\tilde{\alpha}(t)e^{ik_p(\omega_p)z}.
\end{equation}
\noindent The interaction Hamiltonian thus becomes:
\begin{equation}\label{Hinteraction}
\begin{split}
    \hat{H}_\mathrm{I}(t)=A\int_{-L/2}^{+L/2}dz \int d\omega_o \int d\omega_e \hat{a}_o^{(+)}(\omega_o) \hat{a}_e^{(+)}(\omega_e) \\ \tilde{\alpha}(t) e^{-i\{[k_o(\omega_o)+k_e(\omega_e)-k_p(\omega_p)]z-[\omega_o+\omega_e]t\}} +c.c.
    \end{split}
\end{equation}
\noindent where $L$ is the length of the crystal. The time-dependent part of the interaction Hamiltonian is $\tilde{\alpha}(t)e^{i(\omega_o+\omega_e)t}$. The interaction time is determined by the pulse duration, therefore $\hat{H}_{I}=0$ if $t<t_0$. For $t>t_0$:
\begin{equation}\label{Hinteraction2}
\begin{split}
    \int_{t_0}^{t} dt~ \hat{H}_\mathrm{I}(t) = \\
    = A\int_{-\infty}^{+\infty}dt' \int_{-L/2}^{+L/2}dz \int d\omega_o \int d\omega_e \hat{a}_o^{(+)} \hat{a}_e^{(+)} \\ 
    \int d\omega_p\alpha(\omega_p) \\
    e^{-i\{[k_o(\omega_o)+k_e(\omega_e)-k_p(\omega_p)]z-[\omega_o+\omega_e-\omega_p]t\}}+c.c.
    \end{split}
\end{equation}
\noindent where $\alpha(\omega) \xrightarrow{\mathscr{F}} \tilde{\alpha}(t)$ and $\int_{-\infty}^{+\infty}dt'=2\pi\delta(\omega_o+\omega_e-\omega_p)$.
\begin{equation}\label{Hinteraction3}
\begin{split}
    \int_{t_0}^{t} dt~ \hat{H}_\mathrm{I}(t) = \\
    = 2\pi A \int_{-L/2}^{+L/2}dz \int d\omega_o \int d\omega_e \hat{a}_o^{(+)} \hat{a}_e^{(+)} \alpha(\omega_o+\omega_e) \\
    e^{-i\{[k_o(\omega_o)+k_e(\omega_e)-k_p(\omega_p)]z\}}+c.c.
    \end{split}
\end{equation}
\noindent Integrating over the length of the crystal:
\begin{equation}\label{Hinteraction4}
\begin{split}
    \int_{t_0}^{t} dt'~ \hat{H}_\mathrm{I}(t') = \\
    = 2\pi A \int d\omega_o \int d\omega_e \hat{a}_o^{(+)} \hat{a}_e^{(+)} \alpha(\omega_o+\omega_e)\Phi(\omega_o,\omega_e)+c.c.
    \end{split}
\end{equation}
\noindent where:
\begin{equation}\label{Hinteraction5}
\begin{split}
    \Phi(\omega_o,\omega_e)=\frac{sin[k_o(\omega_o)+k_e(\omega_e)-k_p(\omega_o+\omega_e)]L}{[k_o(\omega_o)+k_e(\omega_e)-k_p(\omega_o+\omega_e)]L}
    \end{split}
\end{equation}
\noindent is the phase-matching function. Under the hypothesis of weak interaction:
\begin{equation}\label{Hinteraction6}
\begin{split}
    |\psi_L\rangle=\frac{2\pi A}{i \hbar}\int d\omega_o \int d\omega_e  \alpha(\omega_o+\omega_e)\Phi(\omega_o,\omega_e)|\omega_o\rangle_o|\omega_e\rangle_e
    \end{split}
\end{equation}
\noindent where $|\omega_o\rangle_o$ is a one-photon Fock state, thus:
\begin{equation}\label{Hinteraction7}
\begin{split}
    |\psi_L\rangle=\frac{2\pi A}{i \hbar}\iint F(\omega_o,\omega_e)|\omega_o\rangle_o|\omega_e\rangle_e d\omega_o d\omega_e.
    \end{split}
\end{equation}
\noindent $F(\omega_o,\omega_e)=\alpha(\omega_o+\omega_e)\Phi(\omega_o,\omega_e)$ is the Joint Spectral Amplitude (JSA).

\section{\label{sec:SI2}Supplementary Note 2: Coincidences to Accidentals Ratio (CAR)}
In the experimental scheme used in this manuscript, in addition to the SPDC photons, the SPADs detect a subtantial number of background photons that orginate either from the BBO crystal photoluminescence or other sources in the lab. The background photons that impinge on both the detectors at a relatively high rate can result in accidental conincidence events, which has no quantum optical origin. A crucial part of any coincidence based experiment is to quantify the rate of such accidental events, in particular relating to the rate of real coincidences: the figure of merit is the so called Coincidence to Accidental Ratio (CAR). 

To evaluate the CAR we run an experiment with the HydraHarp in T3 mode and using one of the two SPAD detectors in the trigger input and the other in one channel source. When the first detector fires, HydraHarp opens an acquisition window and eventual events from the second detector are recorded and time tagged with respect to the trigger from the first detector. 

\begin{figure}[ht]
    \centering
    \includegraphics[width=0.8\linewidth]{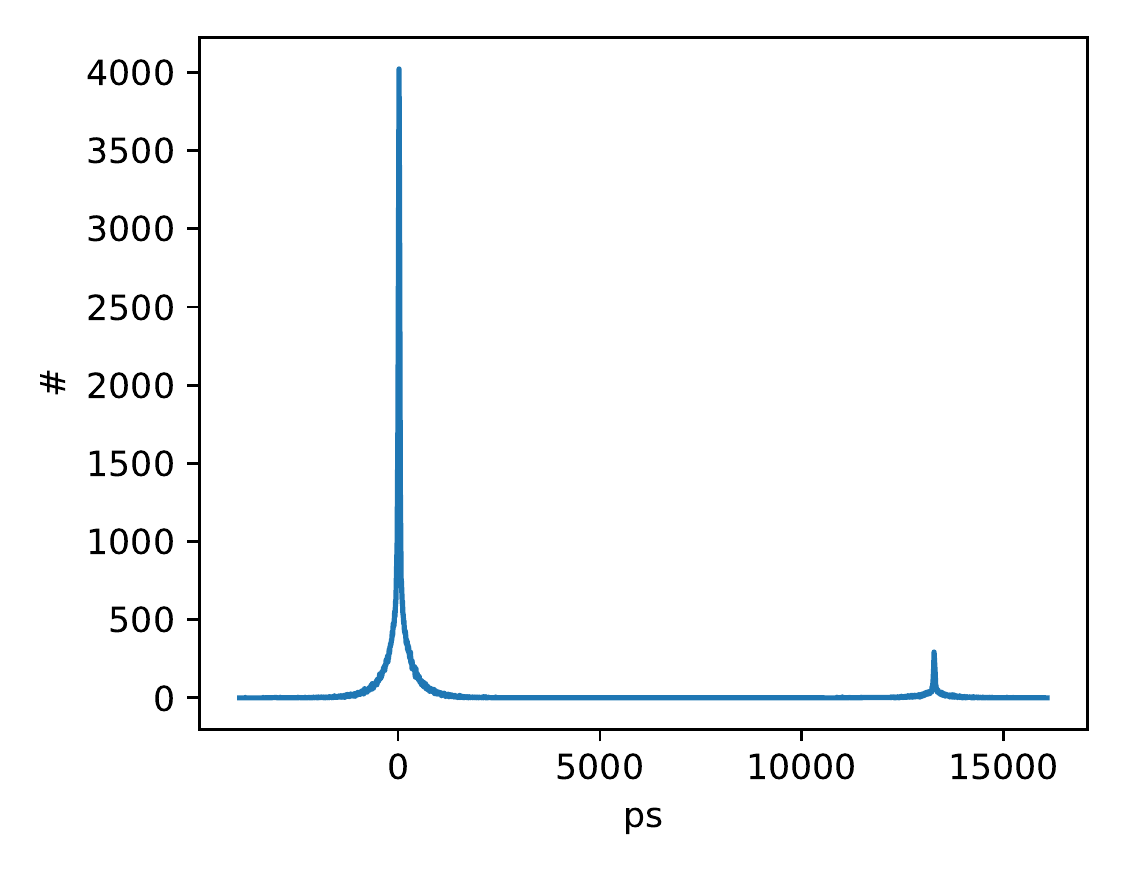}
    \caption{Histogram of time tagged events using one SPAD as trigger source and the other as HydraHarp channel 1. SPDC events are from a BBO Type-I pumped with 10mW.}
    \label{fig_HistoCounts}
\end{figure}

An histogram is built and plotted in Figure \ref{fig_HistoCounts}: it  shows one main peak at time zero and several others evenly spaced by the repetition rate of the laser.

Since SPDC photons pairs are produced, within our experimental resolution, at the same time, they can be located only in the main peak at time zero. Other peaks, on the other hand, are accidental coincidence events. Given that $N_0$ is the number of events in the main peak and $N_1$ is the number of events in the second peak, we compute the CAR metric as following:
\begin{equation}\label{car}
    CAR = \frac{N_{0}-N_1}{N_1}.
\end{equation}
Since accidental events are present also in the main peak, to estimate real coincidence we subtract $N_1$ in the numerator of Eq. \ref{car}. 

\begin{figure}[ht]
    \centering
    \includegraphics[width=\linewidth]{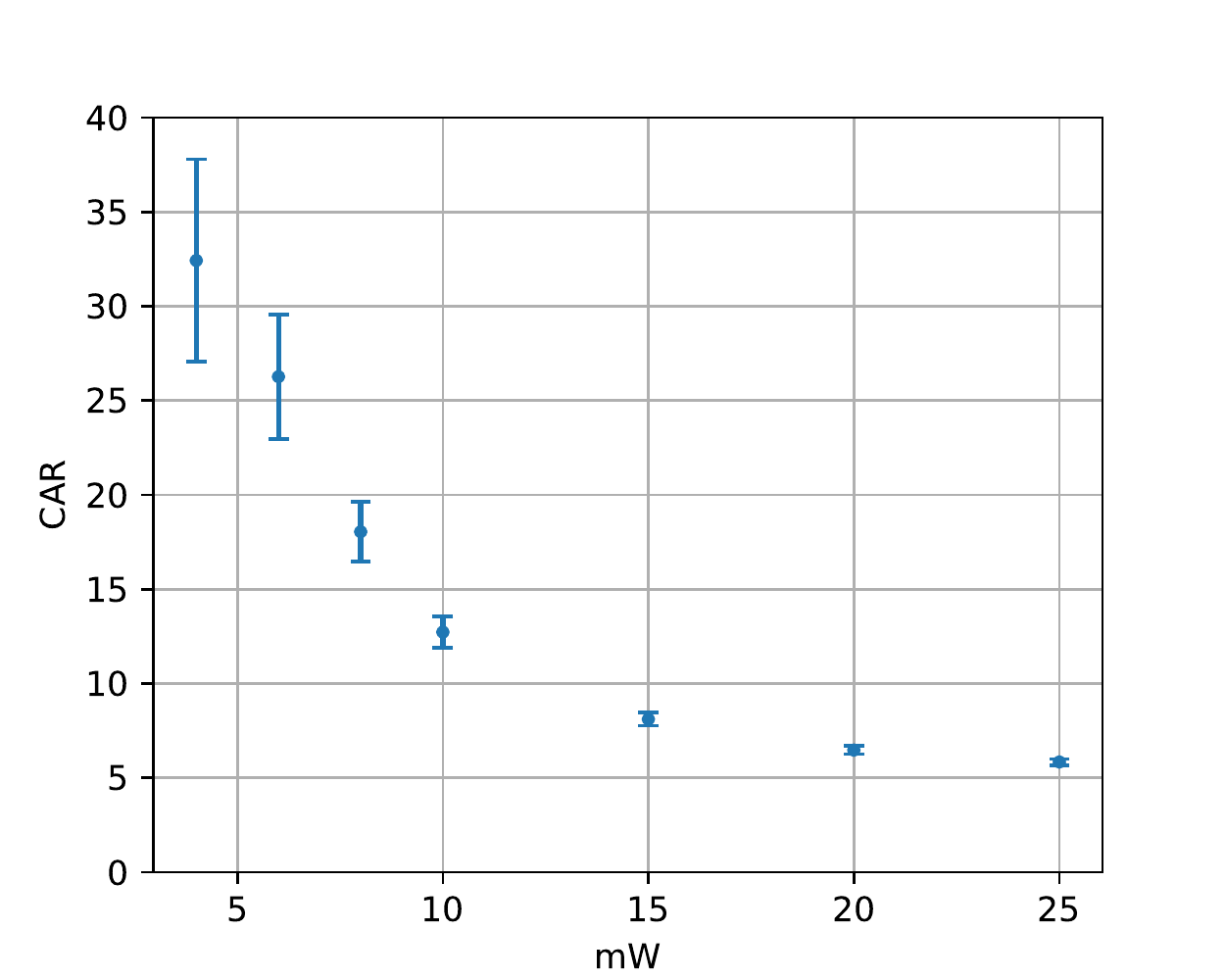}
    \caption{Coincidences to accidentals ratio for several pump powers for type-I crystal.}
    \label{figCAR}
\end{figure}

We evaluate the CAR rate for several pump powers and the result is shown in Figure \ref{figCAR}. Error bars are evaluated considering the counting events $N_0$ and $N_1$ being Poissonian distributed and following propagation of uncertainty for Eq. \ref{car}. JSI shown in section  \ref{subsec:exp_met} has been performed at around 4 mW pump power, with an estimated CAR of $33\pm5$.

\begin{figure}[ht]
    \centering
    \includegraphics[width=\linewidth]{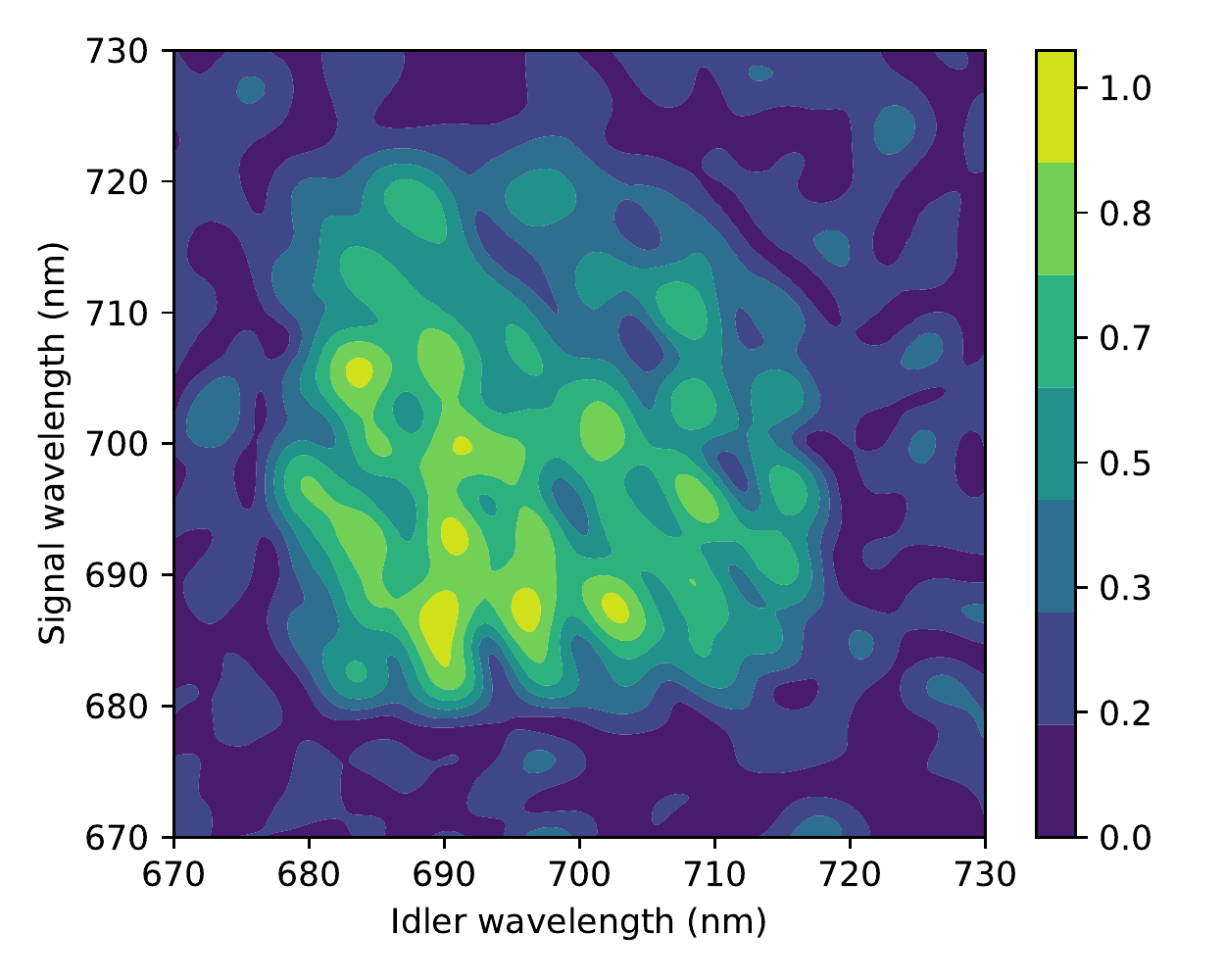}
    \caption{JSI of accidental coincidences with type-I crystal.}
    \label{figJSI_accidental}
\end{figure}

Given our relatively low CAR, we perform an experiment to characterize the spectral signature of the accidental events background. To do so, we run a JSI experiment (with the methodology described in section III A) using as coincidence estimation the second peak from the procedure previously described: in that way we perform a JSI measurement using only accidental events with the goal to characterize the background.  The result is shown in Figure \ref{figJSI_accidental}: the coincidences are an almost uniform background with no spectral correlation limited by the 40nm bandpass filter. We can conclude that the spectral feature of the background events is very distinguishable from the SPDC one and that what we observe in the JSI shown in section III A is mainly due to SPDC photons. 

\section{\label{sec:SI3}Supplementary Note 3: JSA maps trend with $\theta$ and $\alpha$ variation}

\begin{figure*}
    \centering
    \includegraphics[width=.9\linewidth]{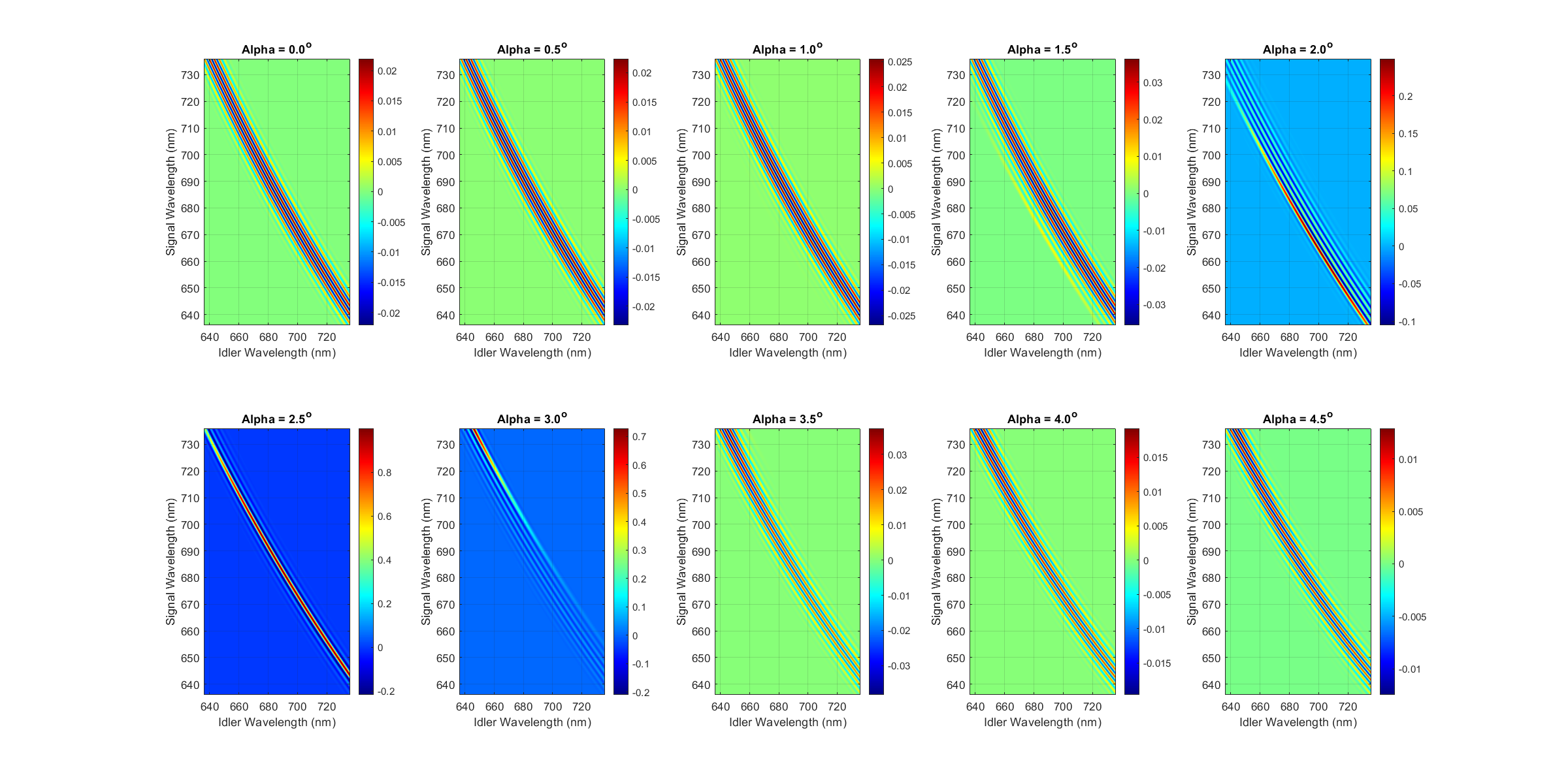}
    \caption{SPDC simulated maps for Non-collinear Type I with $\theta=34.93^o$ and $\alpha$ spanning from $0$ to $4.5^o$.}
    \label{SI_fig_SPDC_type1_alpha}
\end{figure*}

In order to demonstrate what we stated about the Entropy variation with respect to $\theta$ and $\alpha$, we show here, as an example, the JSA maps for the non-collinear Type-I case with a phase matching angle of $\theta = 34.93^o$ and at different $\alpha$ (Fig~\ref{SI_fig_SPDC_type1_alpha}) . For angles between about $2^o$ and $3^o$, the maxima of the phase matching function is still within the range of the pump envelope function, leading to a clean JSA map and large entropy. At angles below and above these values, only the small oscillatory components of the $sinc$ PMF falls within the range and in those cases, the JSA is mostly reminescent of the pump envelope function. This results in an apparent convergence of the entropy values (as seen in Fig~\ref{fig4_AngleDep}). 
\begin{figure}
    \centering
    \includegraphics[width=.9\linewidth]{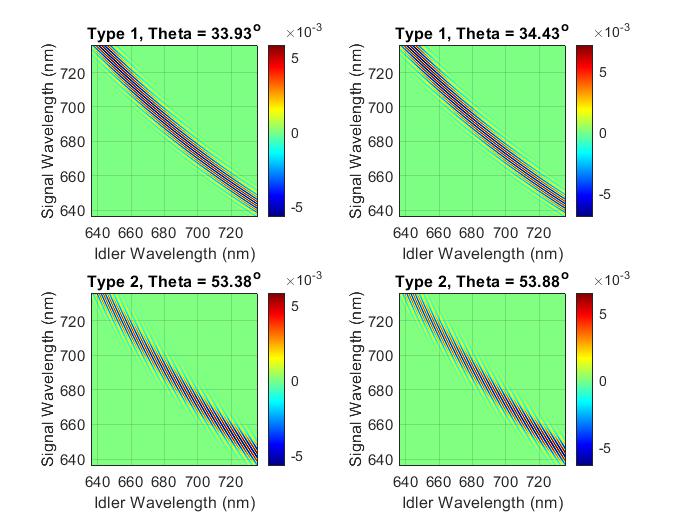}
    \caption{SPDC maps with angle $\alpha=5^o$ for Non-collinear Type I and Type II with 2 different $\theta$ angles.}
    \label{SI_fig_SPDC_angles}
\end{figure}
This can also be seen in other crystal configurations as well, and for both Type-I and Type-II cases. We show in Fig~\ref{SI_fig_SPDC_angles} the JSA simulated maps for Type I and Type II with two different $\theta$ each and with $\alpha=5^o$, where the Entanglement Entropy becomes the same for both the cases. In all of them, the JSA is dominated by the shape of the pump envelope function and only minute contributions from the tails of the phase matching function, thus resulting in the same value for the entropy.


\end{document}